\documentclass[prl,aps,twocolumn,superscriptaddress]{revtex4-2}

\usepackage{latexsym} \usepackage{graphicx}
\usepackage{amssymb} \usepackage{bm} \usepackage{framed}
\usepackage{lipsum} \usepackage{color} \usepackage{amsmath}
\usepackage{footmisc} \usepackage{hyperref}  \usepackage{url}
\usepackage{tikz}

\begin{document}

\title{Magnetic microswimmers exhibit Bose-Einstein-like condensation} 

\author{Fanlong Meng}
\affiliation{Rudolf Peierls center for Theoretical Physics, University of Oxford, Oxford OX1 3PU, UK}
\affiliation{Max Planck Institute for Dynamics and Self-Organization, G\"{o}ttingen 37077, Germany}
\affiliation{CAS Key Laboratory for Theoretical Physics, Institute of Theoretical Physics, Chinese Academy of Sciences, Beijing 100190, China}

\author{Daiki Matsunaga}
\affiliation{Rudolf Peierls center for Theoretical Physics, University of Oxford, Oxford OX1 3PU, UK}
\affiliation{Graduate School of Engineering Science, Osaka University, 5608531 Osaka, Japan}

\author{Beno\^it Mahault}
\affiliation{Max Planck Institute for Dynamics and Self-Organization, G\"{o}ttingen 37077, Germany}

\author{Ramin Golestanian}
\email{ramin.golestanian@ds.mpg.de}
\affiliation{Rudolf Peierls center for Theoretical Physics, University of Oxford, Oxford OX1 3PU, UK}
\affiliation{Max Planck Institute for Dynamics and Self-Organization, G\"{o}ttingen 37077, Germany}


\begin{abstract}
We study an active matter system comprised of magnetic microswimmers confined in a microfluidic channel and show that it exhibits a new type of self-organized behavior. Combining analytical techniques and Brownian dynamics simulations, we demonstrate how the interplay of non-equilibrium activity, external driving, and magnetic interactions leads to the condensation of swimmers at the center of the channel via a non-equilibrium phase transition that is formally akin to Bose-Einstein condensation. We find that the effective dynamics of the microswimmers can be mapped onto a diffusivity-edge problem, and use the mapping to build a generalized thermodynamic framework, which is verified by a parameter-free comparison with our simulations. Our work reveals how driven active matter has the potential to generate exotic classical non-equilibrium phases of matter with traits that are analogous to those observed in quantum systems. 
\end{abstract}

\maketitle

The interplay between non-equilibrium collective dynamics of active matter systems~\cite{Gompper2020,Marchetti2013} and external control provides a wide range of possibilities for new classes of self-organization~\cite{Snezhko2011,Zhou2014,Guillamat2016,Cohen2014,Ellis2017,Waisbord2016,Aubret2018,Vincenti2019,MassanaCid2019,Matsunaga2019,Han2020,Golestanian2019BEC}. 
Due to the versatility it offers~\cite{Erb2009}, magnetic actuation \cite{Vilfan2009,Coq2011,Meng2019} and magnetic steering~\cite{Tierno2008b,Bente2018} has received increasing attention in the recent years, with experimental realizations consisting of both biological~\cite{Blakemore1975,faivre2008} and synthetic~\cite{Dreyfus2005,Tierno2008,Tierno2010,Grosjean2016,Grosjean2018}
microswimmers. 

The coupling between non-equilibrium activity and long-range magnetic dipole-dipole interaction can lead to new emergent properties for magnetic microswimmers~\cite{Rupprecht2016,Waisbord2016,Alonso2018,Meng2018,Vincenti2018,Koessel2019,Vincenti2019}. Similarly rich phenomenology is known to emerge from long-range interactions in phoretic active matter \cite{RG-phoretic}, and in particular, due to the interplay between translational and orientational degrees of freedom \cite{saha2014clusters,Saha_2019,Kranz-trail-PRL:2016,Gelimson-trail-PRL:2016}. In determining the potential for such emergent effects, a key difference between biological and artificial magnetic swimmers is in the strength of their respective interactions. While magnetotactic bacteria carry a typical magnetization of the order of $\sim 10 \, \mathrm{A}\cdot \mathrm{m}^{-1}$~\cite{Blakemore1975,faivre2008}, the magnetization can reach values of up to $\sim 10^{3}\,\mathrm{A}\cdot \mathrm{m}^{-1}$ for swimmers with magnetite~\cite{Yan2015}. The effect of such strong magnetic dipole-dipole interaction on the collective response of magnetic swimmers is still largely unexplored, despite the growing interest in their potential applications for cargo and drug delivery in microscopic environments~\cite{felfoul2016}. Here, we illustrate how strong dipole-dipole interactions affect the collective behavior of magnetic microswimmers confined in a microfluidic channel.

Using Brownian dynamics simulations and a coarse-grained analytical framework, we show that the radial dynamics of microswimmers across the channel is equivalent to that of particles diffusing in an effective potential and presenting a diffusivity-edge~\cite{Golestanian2019BEC}. Consequently, the system is found to exhibit a transition leading to the formation of a condensate at the channel center, which coexists with a surrounding gas. By means of a generalized thermodynamic framework, we characterize the singular behavior of the system and find it to be analogous to the characteristics of Bose-Einstein condensation (BEC) transition. These concrete predictions are moreover quantitatively verified by simulations, without the need of tuning parameters. Finally, our extensive simulations across the entire parameter space allow us to construct a phase diagram, with phase boundaries that show agreement with the simple criteria obtained from our analytical framework.


\begin{figure}[t!]
\centering
\includegraphics[width=\columnwidth]{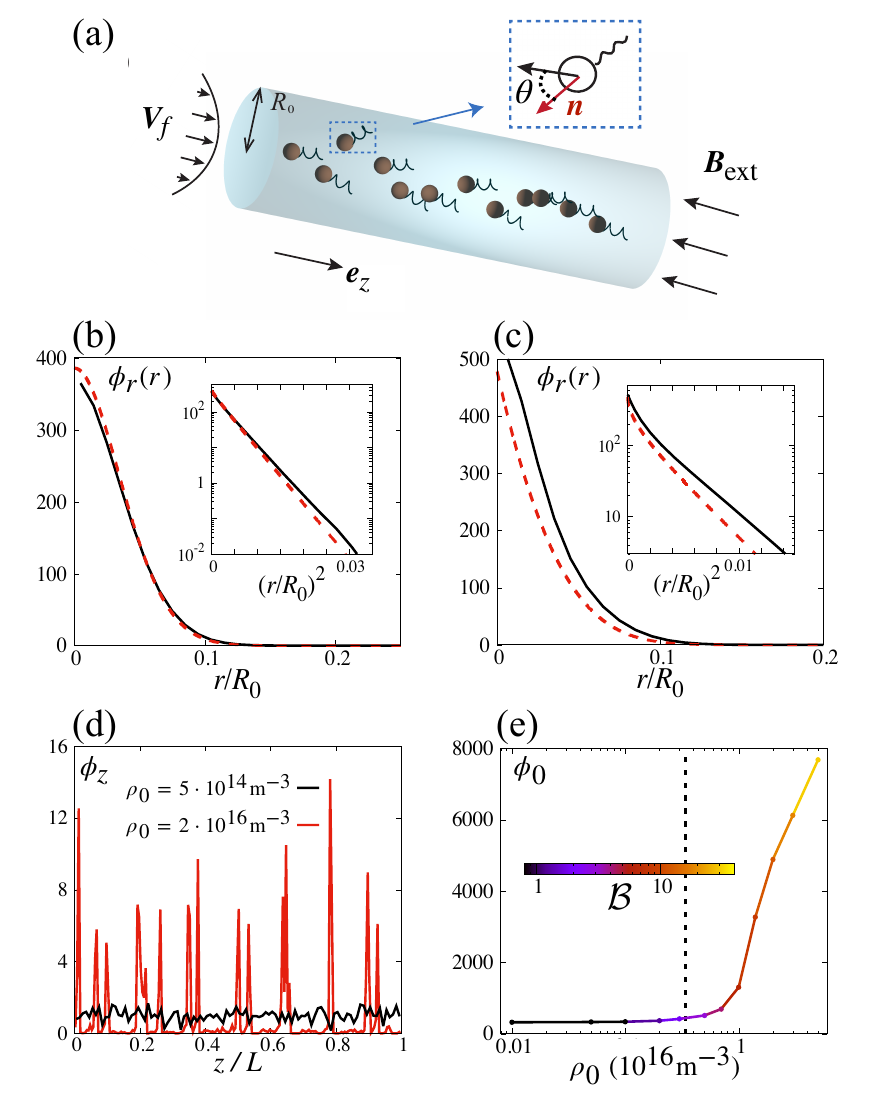}
\caption{{Distribution of the magnetic swimmers in the channel.} (a) Schematics of the system in presence of a Poiseuille flow ${\bm V}_f$ and magnetic field ${\bm B}_{\rm ext}$. Average radial distribution of swimmers $\phi_r$ in the Gaussian (b) and condensed (c) phases at respectively $\rho_0 = 2 \cdot 10^{15}$ ${\rm m}^{-3}$ and $\rho_0 = 5\cdot 10^{15}$ ${\rm m}^{-3}$. Solid (resp. dashed) lines are obtained from simulations (resp. theoretical predictions given by Eqs.~\eqref{eq_phi_sol} and \eqref{eq_phi_sol_cond}). (d) Longitudinal particle distribution of swimmers averaged over $r$ showing self-organization of swimmers into dense clusters for large densities. (e) Maximum value $\phi_0$ of $\phi_r$ at the channel center as function of the mean density $\rho_0$. The vertical dashed line marks the transition to the condensation regime obtained from the theory. In (b)-(e) $v_f = 80$ $\mu{\rm m}\cdot {\rm s}^{-1}$ and $B_{\rm ext} = 4.4$ ${\rm mT}$.
	  }
	  \label{fig1}
\end{figure}


We start by introducing the microscopic model of magnetic microswimmers which we simulated and used as a starting point for the derivation of a field theory. We consider swimmers that carry a magnetic dipole moment $m_0 {\bm n}$, along which they self-propel with a constant speed $v_0$, under the influence of a uniform magnetic field ${\bm B}_{\rm ext} = -B_{\rm ext}\bm{e}_{z}$; see Fig.~\ref{fig1}(a). They swim in a 3 dimensional cylindrical channel oriented along $\bm{e}_{z}$, and experience a Poiseuille flow described as $\bm{V}_{f} = v_{f}(1-r^{2}/R_0^2)\,\bm{e}_{z}$, with $r$ denoting the radial distance from the center and $R_0$ being the channel radius. The Langevin equations governing the dynamics of their position ${\bm r}$ and orientation $\bm n$ thus read
\begin{align}
\dot{\bm r} & = v_0 {\bm n} + \bm{V}_{f} + \frac{m_0}{\zeta} \nabla ({\bm n} \cdot {\bm B}_{\rm int}) + {\bm \xi} \,,
\label{eq_Lagevin_r}\\
\dot{\bm n} & =  \left[ \frac{m_0}{\zeta_r} {\bm n} \times \left( {\bm B}_{\rm ext} + {\bm B}_{\rm int} \right) + \frac{1}{2}\nabla \times \bm{V}_{f}  +  {\bm \xi}_r \right] \times {\bm n} \,, \label{eq_Lagevin_n}
\end{align}
where $\zeta$ and $\zeta_r$ denote the translational and rotational friction coefficients that are taken to be scalar for simplicity, ${\bm \xi}$ and ${\bm \xi}_r$ are thermal noises of respective variances $2 D$ and $2D_{r}$, with $D =  k_{\rm B}T/\zeta$ and $D_r =  k_{\rm B}T/\zeta_r$, and $T$ being the medium temperature. ${\bm B}_{\rm int}$ in Eqs.~(\ref{eq_Lagevin_r}) and (\ref{eq_Lagevin_n}) is the effective magnetic field induced by other swimmers, and is obtained from Amp\`ere's law.
In what follows, $N$ denotes the number of swimmers in the channel, which is set to 1000 (Simulation details can be found in
\footnote{See Supplemental Material at [...] for details about the Brownian dynamics simulations, 
the mean field approximation of the internal magnetic field, the derivation of Eq.~\eqref{condition2}, 
as well as the numerical data leading to Figs.~\ref{fig3}(b) and  \ref{fig3}(c)
}). 
Their mean density $\rho_0 = N/(\pi R_0^2 L)$ is adjusted by varying the channel length $L$. Moreover, we fix $v_0$, $R_0$, $m_0$, $\zeta$, $\zeta_r$, and $T$ to realistic values~\cite{Waisbord2016,Vincenti2019} (see Table I in~\cite{Note1}), such that only $B_{\rm ext}$, $v_f$ and $\rho_0$ are varied. As we shall see later, the dimensionless number
\begin{equation}
{\cal J} \equiv \frac{m_0 B_{\rm ext}}{k_{\rm B} T} \frac{D D_r}{v_0 v_f} \,,
\end{equation}
which combines the relative strength of the magnetic energy versus thermal energy and the strength of propulsion and shear velocities versus diffusion, plays a key role in determining the behavior of the system.

When magnetic interactions are negligible compared to the effect of external driving, the radial dynamics of particles
relaxes over a finite timescale $\tau \equiv D_r R_0^2 m_0 B_{\rm ext}/(v_0 v_f k_{\rm B} T)={\cal J} R_0^2/D$~\cite{Meng2018}.
The dynamics along the channel direction ${\bm e}_z$ is then determined by the value of the dimensionless parameter
\begin{equation}
{\cal B} \equiv \frac{\mu_0 \rho_0 m_0^2}{4 k_{\rm B} T} \frac{m_0 B_{\rm ext}}{k_{\rm B} T} \frac{v_f}{v_0} \,.
\end{equation}
When ${\cal B} < 1$, the distribution of swimmers along ${\bm e}_z$ is uniform on average, whereas for ${\cal B} \ge 1$ the system undergoes an instability leading to a dynamical steady state made of a periodic arrangement of traveling clusters, characterized by strong inhomogeneities in the particle distribution along the channel axis (see Fig.~\ref{fig1}(d)).


In this study, we focus on the stationary radial distribution of swimmers, $\phi_r(r) \equiv \langle \rho({\bm r},t)/\rho_0 \rangle_{z,t}$, 
where $\rho({\bm r},t)$ denotes particle density. The Poiseuille flow ${\bm V}_f$ generates a vorticity that orients the swimmers---that are already aligned by the external magnetic field---towards the center of the channel, essentially acting as a confining potential in the radial direction \cite{Waisbord2016,Meng2018}.
Figure \ref{fig1}(b) shows how for ${\cal B} \ll 1$ $\phi_r(r)$ is well approximated by a Gaussian,
which corresponds to the case where the effective potential is quadratic.
Deep in the clustering phase, $\phi_r(r)$ is not Gaussian and dramatically shoots up in the vicinity of $r=0$ (see Fig.~\ref{fig1}(c)).
The scaling of the maximum of $\phi_r$ at $r=0$, denoted $\phi_0$, with the mean particle density $\rho_0$ is shown in Fig.~\ref{fig1}(e). When ${\cal B}$ is sufficiently small, $\phi_0$ barely varies with $\rho_0$ as expected from a radial focusing by an effective potential. On the contrary, for large values of ${\cal B}$ the system exhibits anomalous accumulation of particles at $r=0$, as indicated by the abrupt increase of $\phi_0$ with $\rho_0$.


The parameter ${\cal B}$ essentially measures how dipole-dipole interactions and alignment with the
external magnetic field dominate over thermal fluctuations, as well as how self-propulsion competes with the external flow.
The condensation phenomenon described above occurs when ${\cal B} \gg 1$,
and thus relies on the key role of magnetic dipolar interactions between swimmers.
As a first simplification, we consider the case where the alignment with $ {\bm B}_{\rm ext} $ dominates over thermal fluctuations:
$m_{0}B_{\rm ext} \gg k_{\rm B}T$.
We moreover assume that the induced magnetic field ${\bm B}_{\rm int}$ is negligible compared to ${\bm B}_{\rm ext}$
in \eqref{eq_Lagevin_n} for the orientational dynamics.

Within the above two assumptions, which are met in most experimentally relevant cases \cite{Waisbord2016,Vincenti2019},
the rotational dynamics becomes a fast process and can be decoupled from the translational dynamics.
The resulting orientational equilibrium is essentially determined by the balance between
magnetic torque ($m_{0}\bm{n}\times \bm{B}_{\mathrm{ext}}$) and the vorticity ($\frac12\bm{\nabla} \times \bm{V}_{f}$).
Denoting $\theta$ the angle between $\bm{n}$ and $-\bm{e}_{z}$ (see Fig.~\ref{fig1}(a)),
its stationary value averaged over thermal fluctuations obeys $\langle\sin\theta \rangle \simeq r / (\tau v_0) \ll 1$.
Projecting \eqref{eq_Lagevin_r} along the radial direction of the channel ${\bm e}_r$,
the equation governing the dynamics of $r$ reads
\begin{equation} \label{eq_ra2}
\dot{r} = - \frac{r}{\tau} - \frac{m_{0}}{\zeta} \partial_{r}\big(\bm{e}_z\cdot \bm{B}_{\mathrm{int}}\big) + \vartheta_{r} \,,
\end{equation}
where the effective Gaussian noise $\vartheta_{r}$ is delta-correlated with variance $2 D_{\rm eff}$, where $D_{\rm eff}\equiv D + v_{0}^2 D_r^{-1} \left[k_{\rm B}T/(m_{0}B_{\mathrm{ext}})\right]^2$.
The dynamics of the system in the radial direction is therefore equivalent to that of interacting dipoles that experience an effective temperature
\begin{equation}\label{eq-Teff}
k_{\rm B} T_{\rm eff} \equiv k_{\rm B} T \left[1 + \frac{v_{0}^2}{D D_r} \left(\frac{k_{\rm B}T}{m_{0}B_{\mathrm{ext}}}\right)^2\right] \,,
\end{equation}
and a confining effective harmonic potential ${\cal U}(r) \equiv \frac12 k r^2$ with stiffness 
$k \equiv \zeta/\tau = k_{\rm B} T_{\rm eff}/(D_{\rm eff}\tau)$
which focuses the particles at the center of the channel. 
Numerical simulations of Eqs.~(\ref{eq_Lagevin_r}) and (\ref{eq_Lagevin_n}) reveal that the clustering instability leads to a highly dynamic regime where clusters assemble and disassemble continuously due to thermal fluctuations (see SM movie~\cite{Note1}, clustering panel). Therefore, longitudinal density inhomogeneities are expected to have little influence on ${\bm B}_{\rm int}$ in the steady state, suggesting that the dynamics will be dominated by the leading contribution of ${\bm B}_{\rm int}(r) \simeq - \mu_0 m_0\rho_0 \phi_r(r) {\bm e}_z$ in this limit~\cite{Note1}.

Inserting this expression into \eqref{eq_ra2}, the radial dynamics completely decouples from the longitudinal one.
Using the following \textit{ansatz} for the particle density inside the channel $\rho({\bm r},t) = \rho_0 \phi_r(r) \phi_z(z,t)$,
the radial equilibrium condition thus follows
\begin{equation} \label{eq_phi_ss}
k_{\rm B} T_{\rm eff} \left( 1 - \frac{\phi_r(r)}{\phi_c} \right)\partial_r \phi_r(r) + \phi_r(r) \partial_r {\cal U}(r) = 0 \,,
\end{equation}
where
\begin{equation}\label{eq-phi_c}
\phi_c \equiv \frac{1}{ 4 {\cal B}}  \frac{v_f^2}{D D_r} \frac{k_{\rm B}T_{\rm eff}}{k R_0^2}=\frac{k_{\rm B}T_{\rm eff} 
}{\mu_0 \rho_0 m_0^2} \,.
\end{equation}
While $T_{\rm eff}$ plays the role of an effective temperature for $\phi_r$ in the dilute limit (corresponding to $\phi_r \to 0$) as mentioned above, the strength of the collective effects leading to density-dependent effective diffusion in \eqref{eq_phi_ss} is set by $\phi_c$. Importantly, we observe that the associated density-dependent effective diffusion coefficient {\it vanishes} at $\phi_r = \phi_c$, which will place the system of magnetic bacteria in a shear flow in the class of systems that can exhibit a classical analogue of Bose-Einstein condensation of particles in the ground state ${\cal U} = 0$ \cite{Golestanian2019BEC,Mahault2020BEC}.

\begin{figure}[t!]
	\centering
 	 \includegraphics[width=\columnwidth]{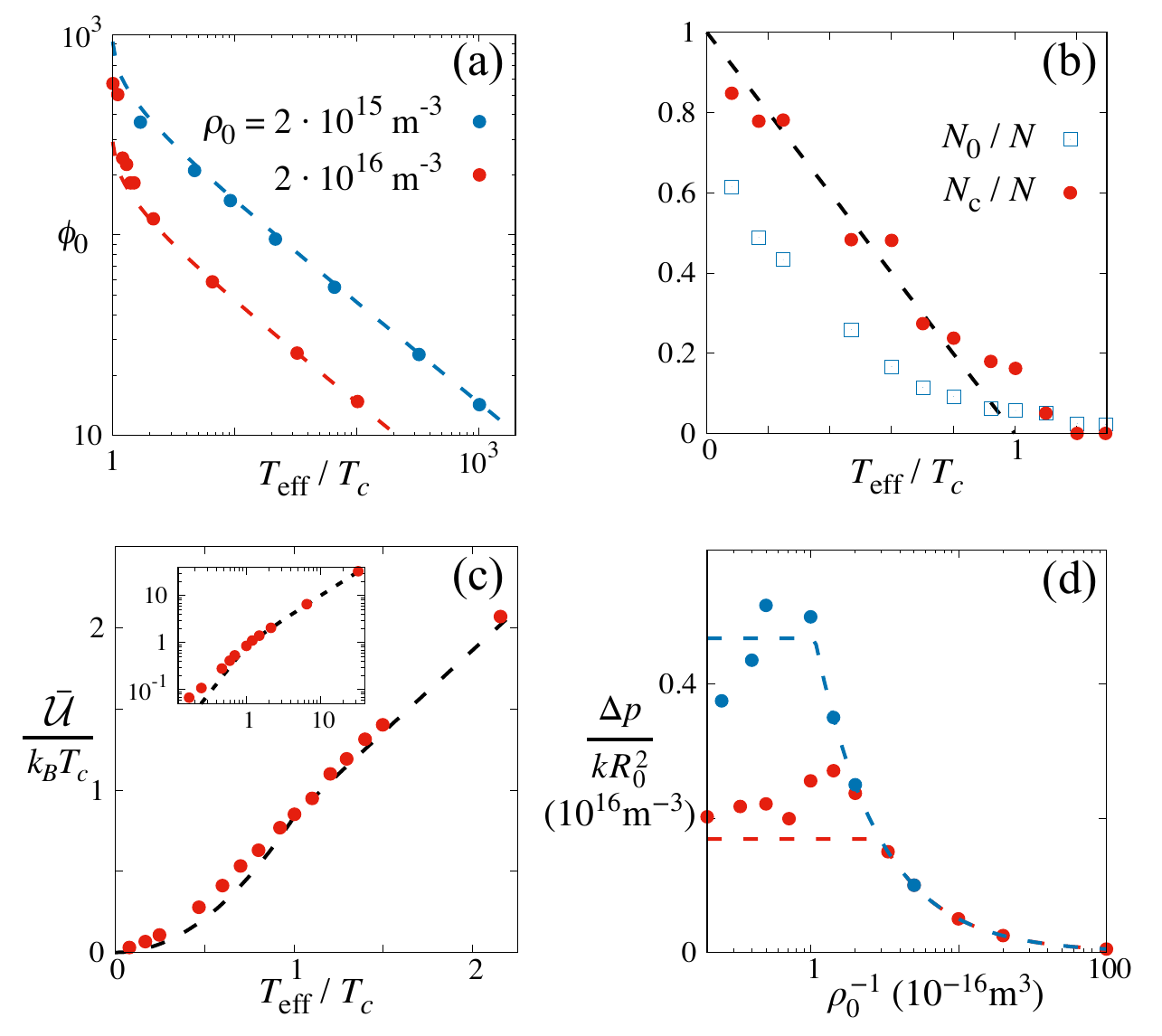}
 	 \caption{Quantitative characterization of the condensate. 
	 (a) Maximum of the radial distribution $\phi_0$, (b) the condensate fraction $N_c/N$, and (c) the mean potential energy per particle $\bar{\cal U}$, as functions of $T_{\rm eff} / T_c$ and fixed mean density $\rho_0$. For (b) and (c) $\rho_0 = 2 \cdot 10^{16}$ ${\rm m}^{-3}$. (d) Pressure exerted on the condensate $\Delta p$ as a function of $\rho_0^{-1}$ for $v_f = 80$ $\mu{\rm m}\cdot {\rm s}^{-1}$, $B_{\rm ext} = 4.4$ ${\rm mT}$ (red) and $v_f = 60$ $\mu{\rm m}\cdot {\rm s}^{-1}$, $B_{\rm ext} = 3.3$ ${\rm mT}$ (blue). 
	 	 In all panels the simulation data (points) are compared to the theoretical predictions (dashed lines) with no free parameters.
	  }
  \label{fig2}
\end{figure}


It follows from \eqref{eq_phi_ss} that for $\phi_r < \phi_c$, $\phi_r$ is a monotonously decreasing function of ${\cal U}$~\cite{Golestanian2019BEC}. We denote $\phi_0$ as the maximum of $\phi_r$ that corresponds to ${\cal U} = 0$, and define $\beta \equiv (k_{\rm B} T_{\rm eff})^{-1}$. Using these definitions, the solution of \eqref{eq_phi_ss} reads
\begin{equation}\label{eq_phi_sol}
	\phi_{r}({\cal U}) \! = \! -\phi_c W_{0}\left[ -\frac{\phi_0}{\phi_c}e^{-\beta{\cal U} - \frac{\phi_0}{\phi_c} }\right] \quad (\phi_0 < \phi_c) \,,
\end{equation}
where $W_{0}(x)$ is the principal branch of the Lambert $W$ function that satisfies $W_{0}\left(xe^x\right) = x$.
When $\phi_0 < \phi_c$, $\phi_{r}$ is a smooth function of $r$ and its normalization
$\frac{2}{R_0^2} \int_0^{\infty} r {\rm d}r \, \phi_r({\cal U}(r)) = \tfrac{2}{k R_0^2} \int_0^{\infty} {\rm d}{\cal U} \, \phi_r({\cal U}) = 1$ leads to \footnote{In all this work, we consider the case where the radial focusing of particles due to the harmonic potential ${\cal U}$ occurs on scales $r \ll R_0$, such that the value of $\phi_r$ at $r=R_0$ can be neglected and all integrals over $r$ are evaluated from $0$ to $+\infty$.}
\begin{equation} \label{eq_phi0}
\phi_0 = \phi_c \left( 1 - \sqrt{1 - T_c / T_{\rm eff}} \right) \quad ( T_{\rm eff} > T_c ) \,,
\end{equation}
where $k_{\rm B} T_c \equiv k R_0^2/\phi_c$ is defined as the value of $k_{\rm B} T_{\rm eff}$ for which $\phi_0 = \phi_c$.
In particular, for $T_{\rm eff} \gg T_c$ the effect of dipolar interactions is negligible,
such that $\phi_r$ is well approximated by a Boltzmann distribution:
$\phi_r \sim \tfrac{\beta k R_0^2}{2} \exp(-\beta {\cal U})$, 
in agreement with our numerical simulations (see Fig.~\ref{fig1}(b)).
As the systematic derivation of \eqref{eq_phi_ss} from the particle-level stochastic dynamics gives
the expressions of $T_{\rm eff}$, $\phi_c$, and ${\cal U}$ as functions of the microscopic parameters,
a quantitative and parameter-free comparison between the theory and the simulations is possible. This is shown in Fig.~\ref{fig2}, whose panel (a) verifies that \eqref{eq_phi0} is in excellent agreement with the simulation results for $T_{\rm eff} > T_c$.

\begin{figure}[t!]
	\centering
 	 \includegraphics[width=\columnwidth]{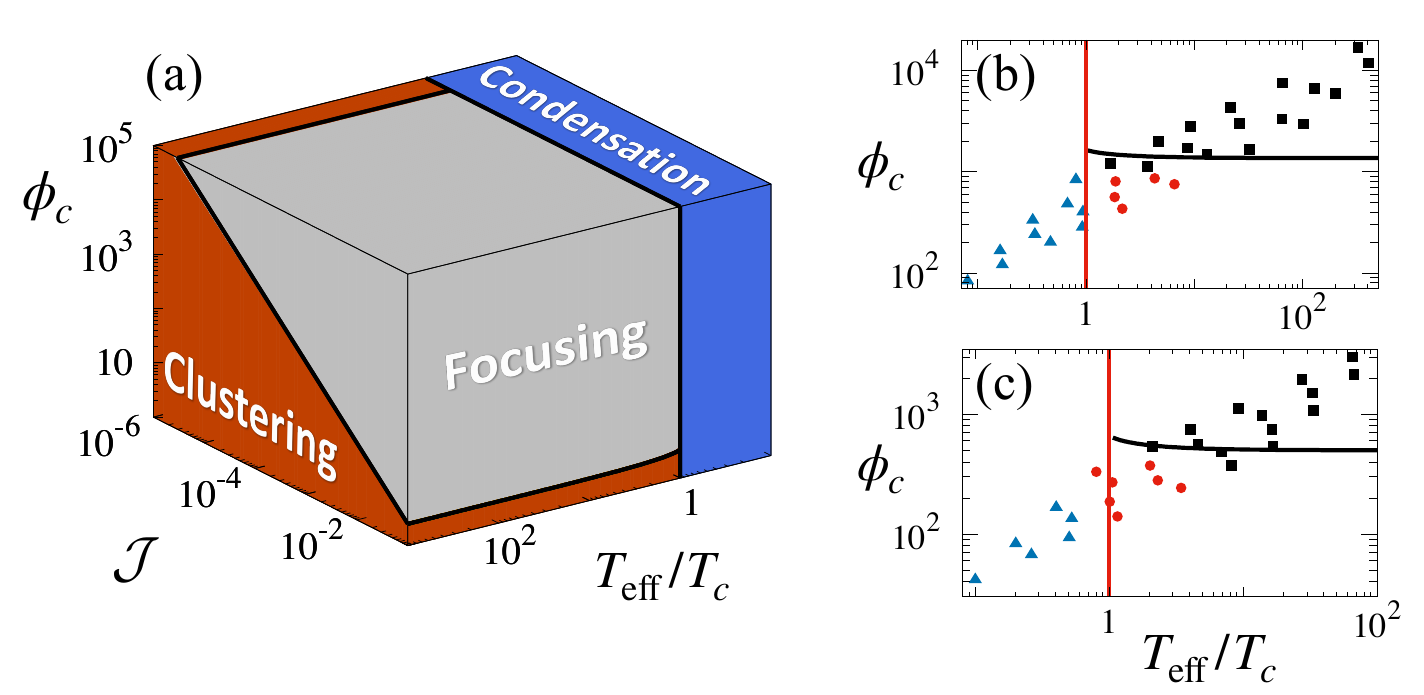}
 	 \caption{{Dynamical regimes of the magnetic swimmers.} (a) Theoretical phase diagram in the ($T_{\rm eff}/T_c$, ${\cal J}$, $\phi_c$) parameter space showing the three different phases defined in the text: focusing, clustering and condensation. 
	  (b) and (c) Comparison of theoretical transition lines to clustering (black) and condensation (red) with Brownian dynamics simulations (black squares: focusing; red circles: clustering; blue triangles: condensation) at fixed values of ${\cal J}=2\cdot 10^{-4}$ (b) and $5\cdot 10^{-4}$ (c) (see~\cite{Note1} for details).
	  }
  \label{fig3}
\end{figure}

			
As $\phi_0$ approaches $\phi_c$ (or equivalently as $T_{\rm eff} \to T_c$), $\phi_r$ becomes non-analytical at ${\cal U} = 0$,
which reflects the formation of a condensate (see Fig.~\ref{fig1}(c)). Consequently, the distribution cannot be normalized and
an additional contribution from the ground state needs to be added by hand. Denoting the number of particles in the condensate as $N_c$ (and $N_c/N$ as the corresponding ground state fraction), the distribution reads
\begin{equation}\label{eq_phi_sol_cond}
    \phi_{r}({\cal U}) \! = \! \frac{k R_0^2}{2}\frac{N_c}{N} \delta({\cal U}) -\phi_c W_{0}\left[ -e^{-\beta{\cal U} - 1 }\right]
    \quad ( T_{\rm eff} \le T_c) \,.
\end{equation}
This is the only stable solution of \eqref{eq_phi_ss} which admits values of $\phi_r$ larger than $\phi_c$~\cite{Mahault2020BEC}.
It thus emerges that the system belongs to the class of diffusivity-edge problems treated in Ref.~\cite{Golestanian2019BEC}, for which the nonlinear diffusion vanishes for all $\phi_r \ge \phi_c$, because the solution given in \eqref{eq_phi_sol_cond} admits values of $\phi$ larger than $\phi_c$ at a single point only.

From the normalization of $\phi_r$, we find that the fraction of particles in the condensate satisfies the following relation
\begin{equation} \label{eq_cond_frac}
\frac{N_c}{N} = 1 - \frac{T_{\rm eff}}{T_c} \quad ( T_{\rm eff} \le T_c ) \,,
\end{equation}
which is consistent with the BEC law in two-dimensional free space~\cite{Ziff1977BoseGas,Note3}.
Equation \eqref{eq_phi_sol_cond} predicts the formation of a point-wise condensate at ${\cal U} = 0$. We have measured $N_c/N$ in our Brownian dynamics simulations by defining it as $\tfrac{2}{k R_0^2}\int_{\phi_r({\cal U}) \ge \phi_c} {\rm d}{\cal U} \phi_r({\cal U})$, and found a good agreement with the theoretical prediction of \eqref{eq_cond_frac} as shown (by the red dots) in Fig.~\ref{fig2}(b). We have found that measuring the average number of particles $N_0$ in a cylinder of radius $0.005 R_0$ around the channel center leads to an underestimation of the number of condensed particles (see the hollow squares in Fig.~\ref{fig2}(b)). We thus conclude from these observations that the condensate emerging in our simulations occupies a finite volume. This feature can moreover be read directly from the distribution $\phi_r$ and is linked to the regularization of the near-field magnetic interactions, whose details are discussed in~\cite{Note1}. The addition of short-range repulsion between the swimmers is expected to have similar consequences.


Following previous works \cite{Golestanian2019BEC,Mahault2020BEC}, the analogy with BEC can be further extended by defining and calculating thermodynamic quantities for the system. The mean potential energy per particle defined as $\bar{\cal U} \equiv \tfrac{2}{k R_0^2}\int_0^{\infty} {\rm d}{\cal U} \, {\cal U}\phi_r({\cal U})$, can be explicitly calculated to give
\begin{equation} \label{eq_u}
\frac{\bar{\cal U}}{k_{\rm B} T_{\rm eff}} = \left\{\begin{array}{ll}
\vspace{1mm}
 \frac{5}{6} + \frac{1}{3} \frac{\phi_0}{\phi_c}\left(\frac{T_{\rm eff}}{T_c} - 1\right) & (T_{\rm eff} > T_c) \\
 \frac{5}{6} \frac{T_{\rm eff}}{T_c} & (T_{\rm eff} \le T_c)
\end{array} \right. \,.
\end{equation}
As in the case of BEC~\footnote{In quantum BEC, the condensation transition only takes place in dimensions $d \ge 3$, 
while the total energy shows a discontinuous slope at $T_c$ for $d \ge 5$~\cite{Ziff1977BoseGas}. In our case, it is possible to have the transition in dimensions $d < 3$, for which the same dependence on dimensionality as in the quantum BEC case is observed. The reason behind this subtle different with the quantum BEC is explained in \cite{Golestanian2019BEC}.}, \eqref{eq_u} predicts a change of slope of $\bar{\cal U}$ at $T_{\rm eff} = T_c$~\cite{Ziff1977BoseGas}.
In particular, for $T_{\rm eff} \gg T_c$ it gives $\bar{\cal U} \sim k_{\rm B} T_{\rm eff}$, which corresponds to a 2-dimensional ideal gas law. Figure~\ref{fig2}(c) shows that the theoretical prediction (\eqref{eq_u}) is well reproduced by the simulations. We note that $\bar{\cal U}$ is generally overestimated in the condensed phase as a consequence of the finite radial extension of the condensate.


With the parameters used in the simulations, the radial focusing of particles occurs on scales much smaller than the channel radius (see Figs.~\ref{fig1}(b) and (c)). Therefore, the radial confinement of particles is essentially due to the effective potential ${\cal U}(r)$
and the effect of the channel boundary is negligible. Within the generalized thermodynamics of the BEC in the steady state, a pressure $p$ can be defined for the active fluid via \cite{Golestanian2019BEC,Mahault2020BEC} ${\rm d}p = -\rho_0\phi_r({\cal U}) {\rm d}{\cal U}$. The difference of pressure between the edge and center of the channel is thus given by
\begin{equation} \label{eq_pressure}
\Delta p = \frac12{\rho_0 k R_0^2} \times \left\{\begin{array}{cc}
\vspace{1mm}
1 & (T_{\rm eff} > T_c) \\
T_{\rm eff}/T_c & (T_{\rm eff} \le T_c)
\end{array} \right. \,,
\end{equation}
which corresponds to the behavior expected for BEC \cite{Ziff1977BoseGas,Note3,BEC-pressure}. In particular, as $T_c \propto \phi_c^{-1} \propto \rho_0$, $\Delta p$ is independent of $\rho_0$ in the condensed phase ($T_{\rm eff} \le T_c$). Isotherms of $\Delta p$ versus $\rho^{-1}_0$ as represented in Fig.~\ref{fig2}(d) therefore exhibit characteristic plateaus for $\rho_0^{-1} \le \rho^{-1}_{0,c}$, where $\rho_{0,c}$ is the value of $\rho_0$ at which $T_c = T_{\rm eff}$. To compare \eqref{eq_pressure} with the simulation, we have calculated the total pressure exerted on the condensate, defined via $\rho_0 \int_{\phi_r({\cal U}) < \phi_c} {\rm d}{\cal U} \phi_r({\cal U})$, by using the measured average distribution $\phi_r$. $\Delta p$ exhibits a good agreement with the theory as seen in Fig.~\ref{fig2}(d). As in the case of mean energy $\bar{\cal U}$, $\Delta p$ appears to be larger than the predicted values in the condensed phase, which is also due to the finite volume occupied by the condensate.


The longitudinal clustering of swimmers was characterized in a previous study~\cite{Meng2018} neglecting
the dipole-dipole interaction term in \eqref{eq_ra2}, thus effectively describing the limit $\phi_0/\phi_c \ll 1$.
As larger values of $\phi_0/\phi_c$ qualitatively change the shape of the radial distribution, a natural question is how these modifications affect the behavior of swimmers in the longitudinal direction.
Using \eqref{eq_phi_sol}, we find that the longitudinal clustering instability occurs for $T_{\rm eff} > T_c$ when the following condition is satisfied~\cite{Note1}
\begin{equation}\label{condition2}
 \frac{1}{{\cal J}\phi_c}\left(\frac{T_{\rm eff}}{T_c}\right)^2 \left(\frac{\phi_0}{\phi_c}\right)^2  \left(1-\frac{2\phi_0}{3\phi_c}\right) \ge 1 \,.
\end{equation}
In the limit of small $\phi_0/\phi_c$ and with parameter values considered in Ref.~\cite{Meng2018},
\eqref{condition2} reduces to ${\cal B} \ge 1$, as expected. Below the condensation threshold where $T_{\rm eff} < T_c$ , the lhs of \eqref{condition2} diverges due to the singularity of $\phi_r$ at ${\cal U} = 0$, and the inequality is always satisfied.

Our theoretical investigations thus predict that magnetic microswimmers in a quasi-one dimensional channel exhibit three types of dynamical behavior. When collective effects are negligible, the swimmers are radially focused due to an effective quadratic potential created by the
interplay between the external flow and the magnetic field, while being uniformly distributed along the channel axis. When $T_{\rm eff} > T_c$ and the inequality~(\ref{condition2}) is satisfied, the system undergoes an instability that gives rise to the formation of clusters that travel along the channel. This longitudinal structure formation persists when $T_{\rm eff} \le T_c$, while in that case a macroscopic number of swimmers form a condensate at the center of the channel in a BEC-like fashion. A phase diagram in the ($T_{\rm eff}/T_c$, ${\cal J}$, $\phi_c$) parameter space summarizing this phase behavior is provided in Fig.~\ref{fig3}(a). Our Brownian dynamics simulations at fixed values of $\mathcal{J}=2\cdot 10^{-4}$ and $5\cdot 10^{-4}$ verify our theoretically predicted phase behavior of the system, as shown in Figs.~\ref{fig3}(b) and (c) (see~\cite{Note1} for simulations details).

To conclude, we have fully characterized the collective behavior of magnetic microswimmers suspended in a microfluidic channel.
We have the system to exhibit a novel type of nonequilibrium condensation transition, which shows striking similarities with Bose-Einstein condensation. These findings not only enrich the broad set of many-body dynamics exhibited by active matter systems, but also provide guidelines for future designs of controllable functional micro-robotic active matter systems with desired emergent properties.

\acknowledgements
This work has received funding from the Horizon 2020 research and innovation programme of the EU under Grant Agreement No. 665440, JSPS KAKENHI Grants No. 20K14649, and the Max Planck Society.

\bibliography{reference}

\end{document}